\documentclass[aps,prb,twocolumn,superscriptaddress]{revtex4-1}
\PassOptionsToPackage{english}{babel}

\usepackage[version=3]{mhchem}

\usepackage{graphicx}
\usepackage{amssymb,amsfonts,amsmath}
\usepackage{color}
\usepackage[normalem]{ulem}
\usepackage{morefloats}
\usepackage{soul}
\usepackage[english]{babel}
\usepackage{multirow}
\usepackage{epstopdf}
\usepackage{placeins}
\newcommand{\revision}{\textcolor{black} }

\newcommand{\beginsupplement}{%
        \setcounter{table}{0}
        \renewcommand{\thetable}{S\arabic{table}}%
        \setcounter{figure}{0}
        \renewcommand{\thefigure}{S\arabic{figure}}%
        \setcounter{equation}{0}
        \renewcommand{\theequation}{S\arabic{equation}}%
     }
     
\begin{document}

\author{Sriteja Mantha}
\email{smantha@uni-mainz.de}
\affiliation{Institut f{\"u}r Physik, Johannes Gutenberg Universit{\"a}t Mainz, Staudingerweg 9, 55128 Mainz, Germany}
\author{Shuanhu Qi}
\email{qishuanhu@buaa.edu.cn}
\affiliation{Institut f{\"u}r Physik, Johannes Gutenberg Universit{\"a}t Mainz, Staudingerweg 9, 55128 Mainz, Germany}
\altaffiliation{Key Laboratory of Bio-inspired Smart Interfacial Science and Technology of Ministry of Education, School of Chemistry, Beihang University, Beijing 100191, China}
\author{Matthias Barz}
\email{barz@uni-mainz.de}
\affiliation{Institut f{\"u}r Organische Chemie, Johannes Gutenberg Universit{\"a}t Mainz, Duesbergweg 10-14, 55128 Mainz, Germany}
\author{Friederike Schmid}
\email{schmidfr@uni-mainz.de}
\affiliation{Institut f{\"u}r Physik, Johannes Gutenberg Universit{\"a}t Mainz, Staudingerweg 9, 55128 Mainz, Germany}

\title{How ill-defined constituents produce well-defined nanoparticles: 
Effect of polymer dispersity on the uniformity of copolymeric micelles} 

\begin{abstract}
We investigate the effect of polymer length dispersity on the properties of
self-assembled micelles in solution by self-consistent field calculations.
Polydispersity stabilizes micelles by raising the free energy barriers of
micelle formation and dissolution. Most importantly, it significantly reduces
the size fluctuations of micelles: Block copolymers of moderate polydispersity
form more uniform particles than their monodisperse counterparts.  We attribute
this to the fact that the packing of the solvophobic monomers in the core can
be optimized if the constituent polymers have different length. 
\end{abstract}

\maketitle

Polymeric nanoparticles are attracting considerable and growing interest
because of their potential in materials science \cite{Bukaro2009,
Ethirajan2010, Nasir2015} and in nanomedicine, e.g., in targeted therapeutic or
diagnostic systems \cite{Ulbrich2016,Tibbitt2016}. One attractive way to
fabricate such nanoparticles is to exploit the self-assembly of amphiphilic
block copolymers in solution.  Depending on the lengths and solubilities of the
copolymer blocks, they can form a variety of interesting morphologies, such as
vesicles, rods, and spherical micelles \cite{Foerster2002, Rodriguez2005}.

In the present article, we consider the micelles, which have a core-shell
structure with a solvophobic core and a solvophilic shell exposed to the
solvent\cite{Leermakers1995,Webber1998,Riess2003}. In principle, the core can
be exploited to load the drug molecules, and efficient mechanisms can be
devised to enhance bioavailability \cite{Lebouille2013,Feng2014,Nakayama2006}.
As long as the drug is stably encapsulated, it is believed that the
distribution of drug-loaded polymeric micelles in the body is determined by the
size and surface properties of polymeric micelles rather than the properties of
the drug molecules\cite{Kataoka2001, Cabral2011, Chang2014}. Hence developing
strategies to control the size distribution of polymeric micelles is important
to improve the efficacy of targeted drug delivery techniques. 

However, all synthetic polymers, including block copolymers, possess some
inherent dispersity in the polymer length due to the nature of polymerization
reaction\cite{Seno2007,Lynd2008, Doncom2017}. Currently, it is not understood
to which extent the polydispersity of block copolymers affects the dispersity
of the formed micelles -- especially since even monodisperse amphiphiles do not
form monodisperse micelles \cite{Nelson1997}.  Polydisperse polymer systems
contain individual polymers with shorter or longer blocks. This provides an
entropic advantage in the self-assembly process, since long chains can fill the
center of domains without having to stretch, whereas short chains adopt
conformations near the interface\cite{Lynd2008, Doncom2017}.  Indeed,
experiments on ABA triblock copolymer melts have indicated that polydispersity
greatly enhances the stability of self-assembled lamellar structures
\cite{Widin2012}. Related observations have been made in theoretical studies of
polymer brushes: They indicate that monodisperse brushes show multicritical
behavior \cite{Qi2016,Romeis2015}, which are associated with large anomalous
chain fluctuations \cite{Skvortsov1997, Merlitz2008, Romeis2013} that disappear
in polydisperse brushes \cite{Qi2016}. Hence it is not {\em a priori} clear
whether polydispersity will enhance or reduce the size fluctuations of
self-assembled micelles.

This question is addressed in  the present article. We use self-consistent field
(SCF) theory to study the structures and free energy landscapes of micelles
that assemble from polydisperse polymer solutions for varying polydispersity
index (\DJ) close to the critical micelle concentration (CMC), i.e., the
point where micelles just begin to form.

Studies of polydispersity effects on self-assembled nanostructures are still
comparatively scarce (for reviews see \cite{Lynd2008, Doncom2017}). Most
studies have considered polydispersity effects on the self-organization in
block copolymer melts \cite{Nguyen1994, Matsushita2003, Sides2004, Noro2005,
Wang2005, Ruzette2006, Torikai2006, Lynd2005, Lynd2007A, Lynd2007, Lynd2008,
Park2008, Matsen2006, Matsen2007, Cooke2006, Oschmann2017}. 
Early theoretical studies \cite{Gao1993, Linse1994} on micellar solutions have
predicted that the CMC decreases with polydispersity \cite{Gao1993} and that
the polymers in the micelles are on average longer than in the surrounding
solution. This was later confirmed by experiments \cite{Khougaz1994,
Hvidt2002}.  The size of vesicles formed by amphiphilic diblock copolymers in
solution was reported to decrease with increasing dispersity of the hydrophilic
block\cite{Terreau2003, Jiang2005, Frans2009}.  In contrast, experimental
studies on micelles made of amphiphilic triblocks with polydisperse inner
(hydrophobic) block have shown that the micelle sizes may increase
significantly with \DJ\ (at comparable average chain length)
\cite{Schmitt2012}, in agreement with theoretical predictions \cite{Linse1994},
and that micelles may even become slightly oblate for high polydispersities
\cite{Schmitt2012}. The latter effect is however small (the aspect ratio is
1:1.4), and since we focus on moderate polydispersities in the present work, we
will assume that micelles are spherically symmetric.


{\em Model and Method.} We study systems of polydisperse amphiphilic diblock
copolymers with solvophobic blocks A (chain fraction f) and solvophilic blocks
B (chain fraction (1-f)) immersed in a solvent S, in the grand canonical
ensemble. Since every copolymer length $N$ defines a separate species, we must
introduce separate chemical potentials $\mu_P(N)$ for each. They are
adjusted such that copolymers {\em in solution} are distributed according to a
Schulz-Zimm distribution \cite{Fredrickson2003} $P_{SZ}(N)$ with average chain
length $N_n=\sum_N \: N \: P_{SZ}(N)$ and polydispersity index \DJ $ =
\sum_N \: N^2 P(N)/N_n^2$. Note that the chain length distribution
in the micelles may differ from $P_{SZ}(N)$.

Polymers are modeled as flexible Gaussian chains\cite{Doi,Fredrickson}.
\revision{ 
The segment interactions are characterized by Flory-Huggins parameters
$\chi_{ij}N_n$ ($i,j=A,B,S$) and are chosen similar to previous
work\cite{He2006} as $\chi_{AB}N_n=10.0$, $\chi_{AS}N_n=17.4$ and
$\chi_{BS}N_n=-0.5$. The system is compressible with compressibility (Helfand)
parameter $\kappa_HN_n = 100$.  } The structure and the free energy of
spherically symmetric micelles are calculated within the SCF theory
\cite{Helfand1975,Schmid1998,Fredrickson}.  The model equations, model
parameters, and further details are given in SI.  The micelle free energy $F_m$
is obtained from the free energy difference of a system containing a micelle
and the corresponding (''bulk'') homogeneous system. The micelle radius is
defined as the radius where the solvophobic density assumes the value $\Phi_A=0.5$.
\revision{ In the following, 
spatial distances are reported in units of radius of gyration ($\bar{R}_g$) of
ideal Gaussian chains with length $N_n$ ($\bar{R}^2_g=\frac{N_nb^2}{6}$),
densities are made dimensionless by dividing them by the bulk segment density
$\rho_0$, and energies are given in units of $\bar{C}k_BT$, where
$\bar{C}=\bar{R}^3_g\rho_o/N_n$, is the Ginzburg parameter which characterizes
the strength of fluctuations in the system \cite{Schmid2011,Fredrickson}. 
In the present article, we consider block copolymers with A- and B-blocks
of equal average molecular weight ($\langle f \rangle = 0.5$),
as in the experimental system of Ref.\ \cite{Schmitt2012}.
}


\begin{figure}[t]
\includegraphics[width=3.3in]{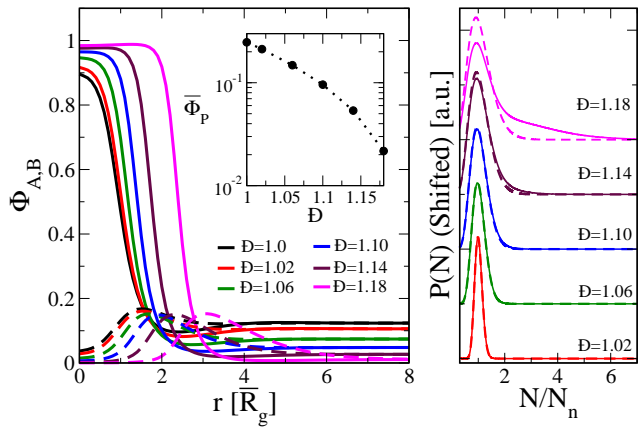}
\caption{Left: Spatial density profiles of solvophobic A segments (full lines)
and solvophilic B segments (dashed lines) in micelles made of A:B copolymers
with f=0.5, $\revision{F_m/k_BT\approx0}$ for different bulk
polydispersity indices \DJ\ as indicated. Inset shows the corresponding shift
of the polymer volume fraction $\bar{\Phi}_p$ in the bulk. Right: Chain length
distribution of copolymers in the bulk (dashed) and in the micelle (solid) for
different values of \DJ. A constant offset has been added for better
visibility.}
\label{fig:fig1}
\end{figure}

{\em Micelle structures and size distributions.} 
\revision{ We first consider copolymers with fixed block ratio $f = 0.5$.}  We
compare systems with same average chain length in the bulk and same micelle
free energy \revision{$F_m/k_B T \approx 0$} (where micelles just begin to
form), but varying polydispersity index \DJ.  To fix $F_m$, the bulk polymer
volume fraction $\bar{\Phi}_p$ must be adjusted. 
\revision{We find that $\bar{\Phi}_P$} decreases significantly with \DJ\ (Fig.\
\ref{fig:fig1}, inset), indicating that micelle formation sets in for smaller
copolymer concentrations in polydisperse solutions. Moreover, the micelle size
increases with \DJ\ (Fig.\ \ref{fig:fig1}, left).  The reason becomes clear
when examining the chain length distribution in the micelles (Fig.\
\ref{fig:fig1}, right). In polydisperse systems, it differs significantly from
the chain length distribution in the bulk.
The largest chains in solution aggregate first, presumably because they can
form aggregates at lower cost of translational entropy, hence micelles become
bigger.  These effects are in agreement with earlier theoretical predictions
\cite{Gao1993,Linse1994} and experimental results \cite{Hvidt2002,
Schmitt2012}.  Similar effects are observed for critical nuclei in polymer
mixtures \cite{Qi2008}. 

\begin{figure}[t]
\includegraphics[width=3.3in]{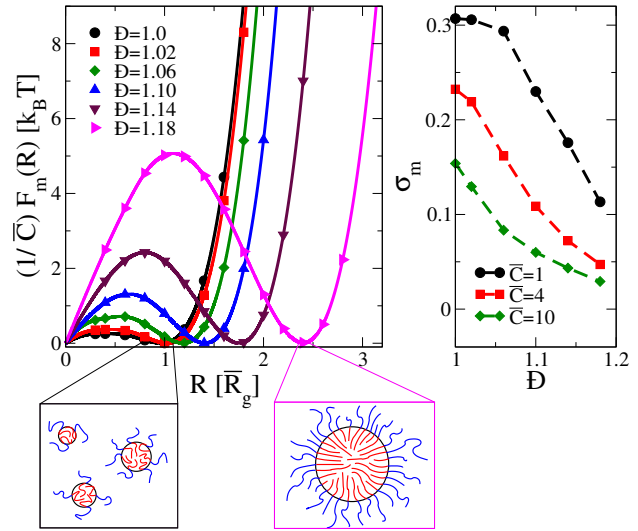}
\caption{Left: Micelle free energy as a function of micelle radius for the same
systems as in Fig.\ \protect\ref{fig:fig1}. Symbols show SCF results, solid
lines a fit to the polynomial $\sum_{n=1}^4 a_n x^n$.  Right:
Corresponding micelle size dispersity $\sigma_m$ as a function of copolymer
polydispersity index \DJ\ for different values of the scaling parameter
\revision{$\bar{C}$}. Cartoons illustrate the proposed stabilizing mechanism: 
Polydisperse solvophobic blocks can pack more efficiently in the core of 
the micelle.
}
\label{fig:fig2}
\end{figure}

Next we study the size distribution of micelles. To this end, we determine the
constrained free energy $F_m(R)$ for micelles of fixed radius $R$ (see SI for
technical details).  The probability for finding a micelle with size $R$ is
then proportional to $\exp(-F_m(R)/k_B T)$.  The function $F_m(R)$ is shown in
Fig.\ \ref{fig:fig2}(left).  It starts at $F_m(0)=0$, then exhibits a maximum
followed by a minimum. The minimum corresponds to the most probable micelle
size $R_{mp}$ and coincides with the solution of the unconstrained SCF
equations discussed above (Fig.\ \ref{fig:fig1}).  Consistent with Fig.\
\ref{fig:fig1}, it shifts to the right with increasing polymer polydispersity.
The maximum correspond to an unstable micelle state:  micelles of this size may
dissolve again. We will refer to it as critical micelle size $R_{mc}$.  The
height of the maximum gives the free energy barrier for micelle formation, and
the free energy difference $\Delta F_m = F_m(R_{mc})- F_m(R_{mp})$ gives the
free energy barrier for micelle dissolution. According to Fig.\ \ref{fig:fig2},
these barriers increase with increasing polydispersity. Hence polydispersity
stabilizes micelles, suggesting that they might also have narrower size
distributions.

The micelle size dispersity is characterized by the relative width of the size
distribution, $\sigma_m = \sqrt{\langle R^2\rangle/\langle R \rangle^2 - 1}$.
To calculate $\sigma_m$, we fit the SCF results for $F_m(R)$ to a fourth order
polynomial (see Fig.\ \ref{fig:fig2}, left) and use that to determine the
averages $\langle R^k \rangle =I_k/I_0$ with $I_k= \int_{R_{mc}}^\infty {\rm
d}R \: R^k \: {\rm e}^{-F_m(R)/k_BT}$.  In doing so, we must specify a value
for the global prefactor \revision{$\bar{C}$} in $F_m(R)$ (see
Fig.\ \ref{fig:fig2}, left).
\revision{ 
The Ginzburg $\bar{C}$ is related to a complementary parameter called the
invariant polymerisation index ($\bar{N}$) as, $\bar{N}\propto\bar{C}^2$, where
$\bar{N}=N_n\rho^2_0b^6$. Typical values for $\bar{N}$ in experimental systems 
are\cite{Glaser2014,Bates1994,Bates1990,Bates1988} $\bar{N}\simeq200-20000$, 
which corresponds to $\bar{C}\simeq1-10$.  }
Fig.\ \ref{fig:fig2} (right) shows our results for $\sigma_m$ as a function of
\DJ\ for three different choices $\revision{\bar{C}} = 1, 4$, and
$10$.  In all cases, the micelle size dispersity $\sigma_m$ decreases with
increasing polymer dispersity \DJ. Thus we find that polymer dispersity not
only stabilizes micelles, but also reduces their size dispersity. This is the
main result of the present article.

\begin{figure}[t]
\includegraphics[width=3.3in]{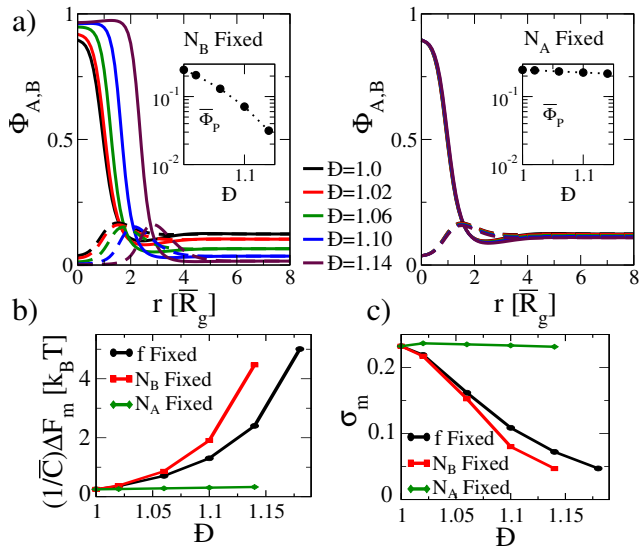}
\caption{
\revision{
(a) Density profiles of A and B segments (full/dashed lines) for micelles with
micelle free energy $\Delta F_m = 0$ in systems where only the solvophobic
(left) or solvophilic (right) chain block is polydisperse.  The inset shows the
corresponding bulk polymer volume fractions $\bar{\Phi}_P$.  
(b) Height $\Delta F_m$ of the free energy barrier for micelle
dissolution.
(c) Width $\sigma_m$ of micelle size distribution at $\bar{C}=4$ vs.\
\DJ\ for all systems considered in the present work. Here \ \DJ\ refers to the
dispersity in the corresponding polydisperse part of the chain (A or B or
total).} }
\label{fig:fig3}
\end{figure}

{\em Other copolymer architectures.} 
To further investigate this phenomenon,
\revision{we study two other classes of systems where the copolymer blocks
still have equal length on average ($\langle f \rangle = 0.5$), but are now
varied independently: In the first system, ths solvophobic block is
polydisperse with ${\left<N_A\right>}/{N_n}=0.5$ and the length of the
solvophilic block is kept fixed at ${N_B}/{N_n}=0.5$. In the second system,
the solvophobic block is fixed at ${N_A}/{N_n}=0.5$ and the length of the
solvopholic block fluctuates with ${\left<N_B\right>}/{N_n}=0.5$.  
}


\revision{ The main results of the SCF calculations are compiled in Fig.\
\ref{fig:fig3}. If only the solvophobic block is polydisperse and the
solvophilic block is kept monodisperse, the effect of polydispersity on the
micelle size (Fig.\ \ref{fig:fig3} a), left), the chain length distribution
(Fig.\ S2 in SI), the height of the free energy barrier $\Delta F_m$ (Fig.\
\ref{fig:fig3} b) and the micelle dispersity (Fig.\ \ref{fig:fig3} c) is even
stronger than before. In contrast, if the solvophobic block is monodisperse,
polydispersity of the solvophilic block has almost no influence on the micelle
size (Fig.\ \ref{fig:fig3}(a), right) and the other micelle characteristics.
Hence the micelle structure and size distribution in the solution is primarily
determined by the dispersity of the solvophobic chain block.} These results
suggest that the main effect of polydispersity is to enhance the packing
efficiency inside the hydrophobic core.

\begin{figure}[t]
\includegraphics[width=3.3in]{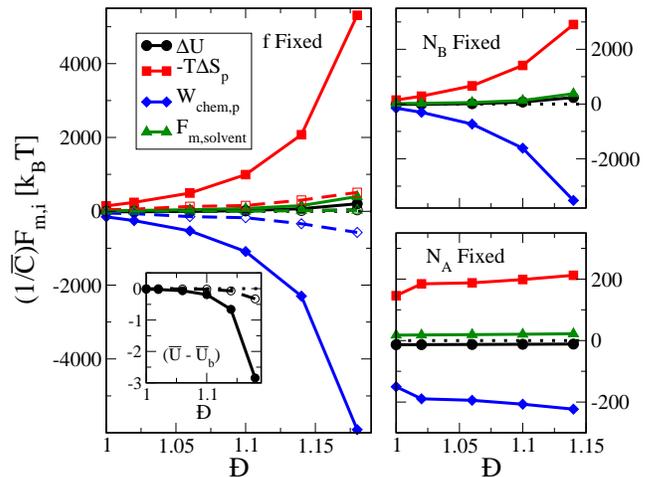}
\caption{Contributions to the micelle free energies from the interaction energy
(black circles), the polymer entropy (red squares), the chemical work
associated with the polymers (blue diamonds) and the sum of solvent entropy
and solvent chemical work (green triangles), for the systems considered in 
this work. Full lines/full symbols refer to most probable micelles, dashed 
lines/open symbols to critical micelles. \revision{Inset shows the 
difference between the internal energy per polymer segment in the micelle 
system and in the bulk (in arbitrary units).}
}
\label{fig:fig4}
\end{figure}

{\em Free energy analysis}.  To test this hypothesis, we separate the different
contributions to the micelle free energy according to $F_m = \Delta
U+W_{\mbox{\tiny chem}} - T \Delta S$, where $\Delta U$ and $\Delta S$ are the
interaction energy and entropy in the micelle relative to the bulk, and
$W_{\mbox{\tiny chem}} = -\mu_s \Delta n_s - \sum_N \mu_p(N) \Delta n_P(N)$
refers to the chemical work required for bringing polymer into the system and
moving solvent out ($\Delta n_x$ is the excess number of molecules of type $X$
in the micelle). Within the SCF framework, the contributions of polymers and
solvent to the entropy and the chemical work can be calculated separately
(see SI for technical details). The results are shown in Fig.\ \ref{fig:fig4}.  
The dominant terms are the polymer entropy and the chemical work associated
with the polymers.

The polymer entropy decreases with increasing polydispersity \DJ.  
This supports the packing hypothesis: Well-packed polymers fluctuate less and
explore fewer conformations, which reduces their entropy.
Interestingly, micelle formation \revision{in polydisperse systems} does not
lead to a reduction of interaction energy $U$: Except in systems with
\revision{fixed solvophobic block length}, $\Delta U$ is positive. The picture
changes in the canonical ensemble, where the number of polymers is fixed and
one must consider the internal energy {\em per polymer segment}. This quantity 
is smaller in micelles than in the bulk 
\revision{ (see Fig.\ \ref{fig:fig4}, inset)}.
Hence micelle formation is driven by a gain in energy per monomer, but not
necessarily by a gain in total energy in a grand canonical setup.  The main
negative contribution to $F_m$ favouring micelle formation is the chemical work
associated with the polymers. It becomes stronger with increasing \DJ.  The
chemical potential ''pushes'' polymers into the solution, and the system gains
free energy if it can accommodate more polymers in the micelles. This again
supports the packing hypothesis. 

The proposed stabilizing mechanism is illustrated in Fig.\ \ref{fig:fig2}
(cartoons). When forming spherical micelle cores from monodisperse solvophobic
blocks, one necessarily creates frustration: Some blocks must stretch and some
must compress to fill the space in the core. Therefore, the micelles offer
little resistance to size variations. In contrast, polydisperse solvophobic
blocks can optimize the packing inside the micelle and minimize the
frustration, which makes them more stable.

{\em Discussion.} In summary, we have investigated the effect of polymer length
dispersity on self-assembled micelles in solutions of amphiphilic diblock
copolymers.  Our main results can be summarized as follows: (i) Consistent with
previous studies \cite{Gao1993, Linse1994, Khougaz1994, Hvidt2002}, we find
that the chain composition in micelles differs from that in solution -- chains
are longer on average. The reason is that in polydisperse systems, long
polymers can segregate from solution and gain energy by forming micelles at
lower cost of translational entropy. As a consequence, the size of the micelles
increases compared to monodisperse systems, in agreement with experimental data
\cite{Schmitt2012}.  (ii) With increasing polydispersity, the free energy
barrier for micelle formation and dissolution increases, and (iii) the width of
the size distribution of micelles decreases.  Hence polymer polydispersity
stabilizes micelles and reduces their size dispersity. A free energy analysis
suggests that this phenomenon is driven by packing in the hydrophobic core,
which is more efficient if the chains are polydisperse.

\revision{In the present work, we have considered a reference ''bulk solution''
where polymer segments are homogeneously distributed in the solution. In
reality, the solvophobic blocks of individual chains may be collapsed
\cite{Wang2010, Wang2012}. This will affect the free energies of the reference
state and the position of the CMC and also have an influence on the chain
length distribution in the micelle.  Unfortunately, studying these effects in a
fully consistent manner is not possible in grand canonical SCF calculations, since
the homogeneous bulk solution serves as outer boundary condition (see, e.g.,article
Fig.\ \ref{fig:fig1}). We plan to analyze this problem in more detail in
the future.}

\revision{We have considered moderate values of the polydispersity up to $\DJ
\approx 1.2$.} For larger values of the polydispersity, it was not possible to
find a solution of the radial SCF equations, suggesting that spherical micelles
may no longer be stable. Indeed, experiments suggest that large
polydispersities may induce shape transitions and even morphological
transitions \cite{Schmitt2012}. This will be an interesting
subject for future work.


We have considered systems close to the CMC, where micelles just begin to form,
such that most copolymers are still in solution. Furthermore, we have assumed
that micelles and micelle size distributions are fully equilibrated. This
corresponds to an experimental situation where micelles are synthesized very
slowly from a solution which does not change with time and provides an
inexhaustible polymer reservoir. In reality, micelles consume polymers and the
polymer composition changes during the process of micelle formation. Moreover,
nanoparticles are not equilibrated. Their sizes, size distributions and even
morphologies depend on the parameters of the synthesis process \cite{Bleul2015,
Nikoubashman2016, Kessler2017}. Nevertheless, we believe that the insights from
the present equilibrium considerations should also be relevant for real
nonequilibrium processes, and could provide useful guidance for experimental
synthesis procedures.  Roughly speaking, our study suggests that it may be
easier to assemble well-defined polymeric nanoparticles with narrow size
distribution from ''bad'' batches of polydisperse building blocks than from
''good'' batches of narrowly distributed building blocks, because the ''bad''
batches provide a range of different molecules which can be combined to
optimize packing. This might be a general principle in solution self-assembly
for nanoparticle synthesis.

\section*{Acknowledgements}

This research was partly supported by the German Science Foundation (DFG) via
SFB 1066  (Grant number 213555243, project Q1) and SFB TRR 146 (Grant number
233630050, project C1). S.Q. acknowledges research support from the National
Natural Science Foundation of China under the Grant NSFC-21873010.  The
simulations were carried out on the high performance computing center MOGON at
JGU Mainz.

\clearpage

\section*{Supplementary Information on: \\
How ill-defined constituents produce well-defined nanoparticles: 
Effect of polymer dispersity on the uniformity of copolymeric micelles}

\beginsupplement
\subsection{Self-consistent field equations for diblock copolymer systems 
with polydisperse polymer chains}

\revision{ To conduct self-consistent field calculations in this work, we
represent the polymer chain length in the units of number average chain
length, $N_n$, and spatial distances in the units of radius of gyration
($\bar{R_g}$) of an ideal Gaussian chain with length $N_n$
($\bar{R}^2_g=N_nb^2/6$). Given these units scale, the grand-canonical free 
energy of a system containing amphiphilic diblock copolymers with number 
average chain length $N_n$ is written as follows \cite{schmid1998}: }

\revision{
\begin{align}
\frac{F}{\bar{C} k_B T} & 
   = \frac{1}{\bar{R}^3_g}\int {\rm d}^3 r \left\{\tilde{\chi}_{AB}\phi_A\phi_B 
         + \tilde{\chi}_{AS}\phi_A\phi_S + \tilde{\chi}_{BS}\phi_B\phi_S \right\} 
    +\frac{\tilde{\kappa}_H}{2\bar{R}^3_g}\int {\rm d}^3 r
     \left( \phi_A + \phi_B + \phi_S -1\right)^2  
   \nonumber \\
   & \qquad 
      -\frac{1}{\bar{R}^3_g}\int {\rm d}^3 r \left\{ \phi_A\tilde{\omega}_A + \phi_B\tilde{\omega}_B 
         + \phi_S\tilde{\omega}_S\right\} 
   \nonumber \\
   & \qquad 
     -\int dX\exp\{\mu_p\left(X\right)/k_B T\} Q_p\left(X\right) 
     -\exp\left\{\mu_S/k_B T\right\}Q_S
\label{eq:FE}
\end{align}
 }
In Equation (\ref{eq:FE}), 
\revision{
$\bar{C}$ is the Ginzburg parameter,
}
$\phi_A, \phi_B$ and $\phi_S$ are rescaled
dimensionless number densities of solvophobic, solvophilic and solvent groups
in the system respectively 
\revision{($\phi_i =  \rho_i/\rho_0$, with $\rho_0$ being bulk segment density),}
\revision{$X$ is polymer segment length (in reduced units, i.e, $X=\frac{N}{N_n}$)},
and \revision{$\tilde{\omega}_A, \tilde{\omega}_B$} and
\revision{$\tilde{\omega}_S$} are corresponding auxiliary fields.  The Helfand parameter
\revision{$\tilde{\kappa}_H$} is an inverse compressibility which is used to keep the local
density of the system almost at a constant value, and \revision{$\tilde{\chi}_{ij}=\chi_{ij}N_n$ are rescaled} Flory-Huggins interaction parameters.

$\mu_S$ is the solvent chemical potential \revision{and $Q_S=\frac{1}{\alpha \bar{R}^3_g} \int {\rm d}^3 r \exp(-\alpha\tilde{\omega}_S({\bf r}))$} is the solvent partition function, \revision{with $\alpha$ being volume occupied by a  solvent molecule relative to that occupied by a polymer molecule with chain length $N_n$ in the solution}. Similarly,
\revision{$Q_p\left(X\right)$} is the single chain partition function and
\revision{$\mu_P\left(X\right)$} is the chemical potential of a chain with \revision{$X$} segments.
Polymers are modelled as Gaussian chains with statistical segment
length $b$.

Chemical potentials in the above equation are computed as follows, using the
method prescribed by Qi et al.\ \cite{Qi2008},
\revision{
\begin{eqnarray}
\exp\{\mu_P\left(X\right)/k_B T\}  
 &=&
    \bar{\phi}_{P}P_{SZ}\left(X\right)\exp\{\tilde{\omega}_{0P}X\} 
\label{eq:muP}
\\
\exp\left\{\mu_S/k_B T\right\}
&=&\bar{\phi}_{S}\exp\left\{\alpha\tilde{\omega}_{0S}\right\}
\label{eq:muS}
\end{eqnarray}
}

In equations \ref{eq:muP} and \ref{eq:muS}, $\bar{\phi}_{P}$ and
$\bar{\phi}_{S}$ are rescaled polymer segment and solvent densities in the bulk
solution that is in equilibrium with the micellar system, and
$\tilde{\omega}_{0P}$ and  $\tilde{\omega}_{0S}$ are corresponding auxiliary
fields. 
\revision{$P_{SZ}(X)$} is the probability of finding a polymer chain with
\revision{$X$} segments in the bulk solution, which is taken to be of
Schulz-Zimm type\cite{Fredrickson2003}:

\revision{
\begin{align}
P_{SZ}(N) = \frac{k^kX^{k-1}}{\Gamma(k)}\exp\left[-kX\right] 
\label{eq:SZ}
\end{align}
}

The polymer polydispersity in the distribution (\ref{eq:SZ}) is determined by
the parameter $k$ and given by \DJ $= 1 + \frac{1}{k}$.
\clearpage
The SCF equations are obtained by a variational extremization
of the free energy with respect to $\phi_A$, $\phi_B$, $\phi_S$, $\tilde{\omega}_A$,
$\tilde{\omega}_B$ and $\tilde{\omega}_S$:

\revision{
\begin{align}
\tilde{\omega}_A({\bf r}) &= \tilde{\chi}_{AB}\phi_B + \tilde{\chi}_{AS}\phi_S 
      + \tilde{\kappa}_H\left( \phi_A + \phi_B + \phi_S - 1 \right) \nonumber \\
\tilde{\omega}_B({\bf r}) &= \tilde{\chi}_{AB}\phi_A + \tilde{\chi}_{BS}\phi_S 
     + \tilde{\kappa}_H\left( \phi_A + \phi_B + \phi_S - 1 \right) \nonumber \\
\tilde{\omega}_S({\bf r}) &= \tilde{\chi}_{AS}\phi_A + \tilde{\chi}_{BS}\phi_B 
     + \tilde{\kappa}_H\left( \phi_A + \phi_B + \phi_S - 1 \right) \nonumber \\
\phi_A({\bf r}) &= \int dX \exp\{\mu_p\left(X\right)/k_B T\} 
     \int_0^{X_A} ds q_f({\bf r},s) q_b({\bf r},X-s) \nonumber \\
\phi_B({\bf r}) &= \int dX \exp\{\mu_p\left(X\right)/k_B T\} 
      \int_0^{X_B} ds q_b({\bf r},s) q_f({\bf r},X-s) \nonumber \\ 
\phi_S({\bf r}) &= \exp\left\{\mu_S/k_B T\right\} \exp\left\{-\alpha\tilde{\omega}_S\right\}
\label{eq:SCF}
\end{align}
}
In Equation (\ref{eq:SCF}), $q_f(r,s)$ and $q_b(r,s)$ are end integrated
forward and backward chain operators, which are obtained from solving the
modified diffusion equation 

\revision{
\begin{equation}
\frac{\partial q({\bf r},s)}{\partial s} 
   = \bar{R}^2_g\nabla^2 q({\bf r},s) 
      - \tilde{\omega}(r) q({\bf r},s),
\label{eq:diff}
\end{equation}
}

with \revision{$\tilde{\omega}({\bf r}) = \tilde{\omega}_A({\bf r})$ for $s < X_A$ in $q_f$ and $s > X -
X_A$ in $q_b$, and $\tilde{\omega}({\bf r}) = \tilde{\omega}_B({\bf r})$ otherwise.} The
Equation \ref{eq:diff} is solved with the initial condition condition
$q_{f,b}({\bf r} ,0)=1$. The single chain partition function for
polymers of length $X$ is given by 
\revision{
\begin{equation}
Q_P(N) = \frac{1}{\bar{R}^3_g} \int {\bf d}^3 r \: q_b({\bf r},X)
       = \frac{1}{\bar{R}^3_g} \int {\bf d}^3 r \: q_f({\bf r},X).
\end{equation}
}

The SCF equations are solved in spherical coordinates using finite
differences and von Neumann boundary conditions.

To calculate the micelle free energy $F_m(R)$ as a function of micelle radius
$R$, we must constrain the radius to a fixed value $R$. This is done by solving
modified SCF equations, which are derived from a modified SCF free energy
functional similar to Equation (\ref{eq:FE}) that however contains a 
virtual additional harmonic term \revision{$k_m (\Phi_A(R)-0.5)^2$ with $k_m= 1.0\times10^{-1}$} (This virtual term is of course omitted when 
calculating $F_m(R)$.)

\subsection{Model parameters used in this work}

The interaction parameters of the model are mostly taken from recent work of He
et al \cite{He2006,He2008}, who investigated vesicle formation in diblock
copolymer solutions using dynamic density functional theory. Starting from the
parameters in their work, we varied $\tilde{\chi}_{AS}$ and $\tilde{\chi}_{BS}$ until stable
spherical micelles were obtained for monodisperse systems. This procedure resulted in the set of parameters 
\revision{
$\tilde{\chi}_{AB}=10.0$, $\tilde{\chi}_{AS}=17.4$, $\tilde{\chi}_{BS}=-0.5$ and we chose $\tilde{\kappa_H}=100.0$  to control the compressibility of the system}
These parameters were then used to investigate the effect of polydispersity 
on the micelles.
\revision{
For numerical reasons we conduct our SCF calculations with $\alpha$ set to 0.1.
}

\begin{figure}[t]
\includegraphics[width=3.5in]{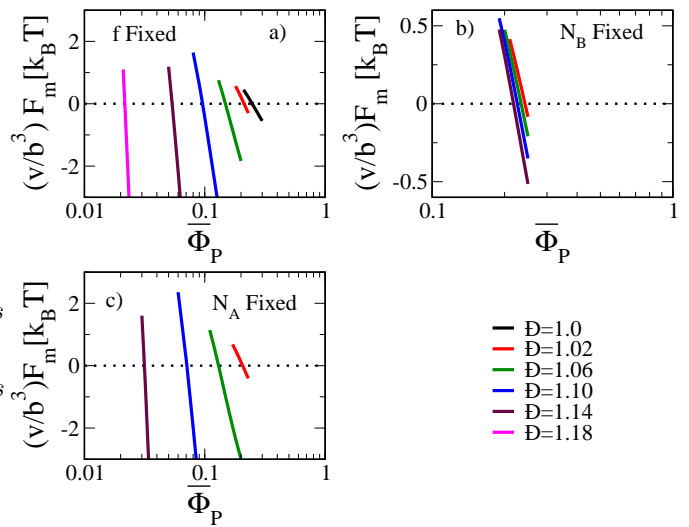}
\caption{Micelle formation energies $F_m$ as a function of bulk polymer 
volume fraction $\bar{\phi}_{p}$ in polymer systems with  (a) $f Fixed$, (b) $N_B$ {\it{Fixed}} and (c) $N_A$ {\it{Fixed.}}
$\DJ$ refers to dispersity in the corresponding polydisperse part of the chain (A or B or total).}

\label{fig:figS1}
\end{figure}

For a copolymeric system with given $f$ and \DJ, spherical micelles are
formed in the system for a range of bulk polymer volume fractions
$\left(\bar{\Phi}_{p}\right)$. In Figure \ref{fig:figS1}, we report micelle
free energies $F_m$ of different systems investigated in this work. As noted in
the main manuscript, $F_m$ is computed as the free energy difference of a
system containing a micelle and the corresponding homogeneous bulk system.
Irrespective of the system investigated, $F_m$ is seen to be a decreasing
function of ($\bar\Phi_{p}$). For given value of $f$, we fix a slightly
positive value of $F_m$. Then, for each \DJ, we choose that particular
$\bar{\Phi}_{P}$ which results in the specified micelle free energy $F_m$.
The model parameters used in this work are summarized in table
\ref{Tab:Param}. 

\clearpage
\begin{table}
\centering
\caption{Parameters to simulate spherical micelles formed by diblock copolymer micelles. $\DJ$ refers to the dispersity in the corresponding polydisperse part of the chain (A or B or total).}
\label{Tab:Param}
\begin{tabular}{|c|c|c|c|c|}\hline
Polymeric system & 
Micelle free energy & k $\left(\mbox{\DJ}\right)$ & $\bar{\Phi}_{p}$ \\ \hline
\multirow{4}{*}{\shortstack{$f Fixed$}}
&\multirow{4}{*}{\shortstack{$F_m\approx3\times10^{-3}\bar{C} k_B T$}}
& 5.5 (1.18) & 2.16e-02 \\ \cline{3-4}
&&7.14 (1.14) & 5.35e-02 \\ \cline{3-4}
&&10 (1.10) & 9.58e-02 \\  \cline{3-4}
&&16.6 (1.06) & 1.48e-01 \\  \cline{3-4}
&&50 (1.02) & 2.12e-01 \\  \cline{3-4}
&&MD (1.00) & 2.47e-01 \\ \hline
\multirow{4}{*}{\shortstack{$N_B$ {\it{Fixed}} }}
&\multirow{4}{*}{\shortstack{$F_m\approx3\times10^{-3}\bar{C} k_B T$} }
& 7.14 (1.14) & 3.15e-02 \\ \cline{3-4}
&&10 (1.10) & 7.10e-02  \\ \cline{3-4}
&&16.6 (1.06) & 1.30e-02 \\ \cline{3-4}
&&50 (1.02) & 2.06e-02  \\  \hline
\multirow{4}{*}{\shortstack{$N_A$ {\it{Fixed}} }}
&\multirow{4}{*}{\shortstack{ $F_m\approx3\times10^{-3}\bar{C} k_B T$}} 
& 7.14 (1.14) & 2.18e-01 \\ \cline{3-4}
&&10 (1.10) & 2.26e-01 \\ \cline{3-4}
&&16.6 (1.06) & 2.34e-01 \\  \cline{3-4}
&&50 (1.02) & 2.43e-01 \\  \hline
\end{tabular}
\end{table}

\subsection{Energy decomposition}

In order to analyze the driving force for the effect of polymer dispersity on
the size distribution of micelles, we decomposed micelle free energy $F_m$ into
the interaction energy ($\Delta U$), the entropy ($T\Delta S$), and the
chemical work ($W_{Chem}$). Within self consistent field theory, the
entropy and the chemical work can be further decomposed into the corresponding
contributions from polymer and solvent components: 

\begin{align}
F_m & = \Delta U - T\Delta S + W_{Chem} \nonumber \\
    &= \Delta U - \left(T\Delta S_P + T\Delta S_S \right) 
       + \left(W_P + W_S\right) 
\end{align}

In this section, we describe how this decomposition is made.
The thermodynamic equation governing grand canonical ensemble is given by

\begin{equation}
\Omega = U - TS - \sum_Z\mu_Z n_Z
\label{eq:GC}
\end{equation}

In the above equation, $\Omega$ is the grand potential, $U$ is the internal
energy, $S$ is the entropy, $T$ is the temperature, $\mu_Z$ is the chemical
potential of species $Z$ and $n_Z$ is the number of particles of type X in the
given system. Note that every copolymer length $X=N/N_n$ defines a separate 
species, hence $Z$ runs over $S$ (solvent) and ($p,X$) (copolymers of length
$X$).  Comparing the Equation (\ref{eq:GC}) to the Equation
(\ref{eq:FE}), the following relations can be deduced:
\revision{
\begin{align}
\frac{U}{\bar{C}k_B T} &= \frac{1}{\bar{R}^3_g}\int {\rm d}^3 r \left\{\tilde{\chi}_{AB}\phi_A\phi_B 
    + \tilde{\chi}_{AS}\phi_A\phi_S +\tilde{ \chi}_{BS}\phi_B\phi_S \right\} \nonumber \\
& \qquad
    + \frac{\tilde{\kappa}_H}{2\bar{R}^3_g}
     \int {\rm d}^3 r \left( \phi_A + \phi_B + \phi_S -1\right)^2 
\label{eq:U}
\end{align}
\clearpage
\begin{align}
-\left(TS + \sum_Z\mu_Z n_Z\right)/\left(\bar{C}k_B T\right)
  = -\frac{1}{\bar{R}^3_g}\int {\rm d}^3 r \left\{ \phi_A\tilde{\omega}_A 
    + \phi_B\tilde{\omega}_B + \phi_S\tilde{\omega}_S\right\} \nonumber \\
    - \int dZ \exp\{\mu_p\left(X\right)/k_B T\} Q_p\left(X\right) 
    - \exp\left\{\mu_S/k_B T\right\}Q_S
\label{eq:Tmu}
\end{align}
}

Equation (\ref{eq:Tmu}) can be readily decomposed into 
polymer (P) and solvent (S) contributions as follows,

\revision{
\begin{align}
-\left(TS + \sum_Z\mu_Z n_Z \right)_{P} / \left(\bar{C}k_B T\right)
  &= - \frac{1}{\bar{R}^3_g}\int {\rm d}^3 r \left\{ \phi_A\tilde{\omega}_A 
          + \phi_B\tilde{\omega}_B\right\}  \nonumber \\
  & \qquad
      -  \int dX \exp\{\mu_p\left(X\right)/k_B T\} Q_p\left(X\right)
\\
-\left(TS + \sum_Z\mu_Z n_Z\right)_{S} /\left(\bar{C}k_B T\right)
  & = -\frac{1}{\bar{R}^3_g}\int {\rm d}^3 r \left\{\phi_S\tilde{\omega}_S\right\} 
     -  \exp\left\{\mu_S/k_B T\right\}Q_S
\end{align}
}

Using Equation (\ref{eq:muP}), the specific contribution 
of polymer molecules to the chemical work can be computed as

\revision{
\begin{align}
W_P & = -\sum_X \mu_P\left(X\right) n_X \nonumber \\
n_N &= \frac{1}{X\bar{R}^3_g} 
    \int {\rm d}^3 r \left(\phi_{A,X} + \phi_{B,X}\right)
\end{align}
where $\phi_{A,X}$ and $\phi_{B,X}$ correspond to the rescaled segment
densities of type A and B of polymer molecules with $X=N/N_n$ segments per chain.
Similarly, using Equation (\ref{eq:muS}), the contribution of solvent particles
to the chemical work can be computed as 
}

\revision{
\begin{align}
W_S &= -\mu_S n_S \nonumber \\
n_S &= \frac{1}{\bar{R}^3_g}\int {\rm d}^3 r \phi_S
\end{align}
}

Based on this decomposition of chemical work into polymer and solvent 
contributions, we can write the corresponding entropic contributions as

\begin{align}
-\left(TS\right)_{P} = -\left(TS + \sum_Z\mu_Z n_Z \right)_{P} - W_P \\
-\left(TS\right)_{S} =- \left(TS + \sum_Z\mu_Z n_Z\right)_{S} - W_S
\end{align}

\clearpage
\subsection{Additional data}

\begin{figure}[t]
\includegraphics[width=3in]{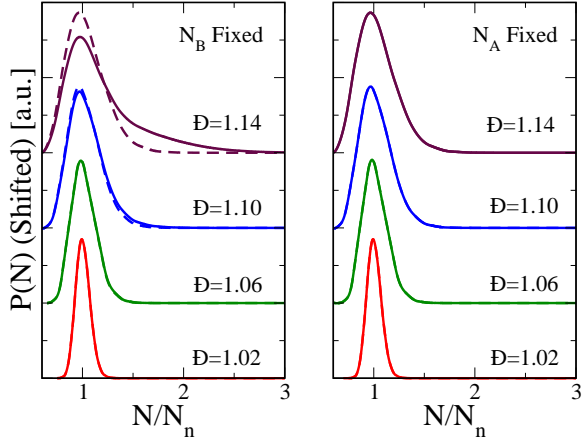}
\caption{Chain length distribution of copolymers in the bulk (dashed) and in the most probable micelle (solid) for (left) systems with $"N_B fixed"$ and (right) for systems with $"N_A fixed"$. Here, $\DJ$ refers to dispersity in the corresponding polydisperse part of the chain (A or B)}
\label{fig:figS2}
\end{figure}

\begin{figure}[t]
\includegraphics[width=3in]{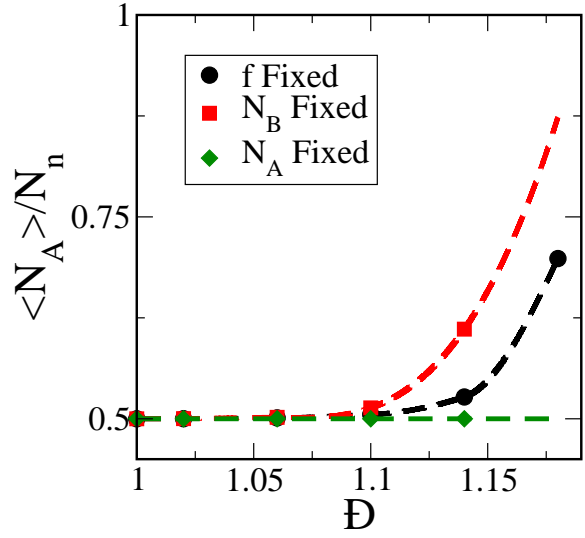}
\caption{Average length of the solvophobic part of the polymer chain in the micelle for systems with different chain architectures at different values of $\DJ$. Here, $\DJ$ refers to dispersity in the corresponding polydisperse part of the chain (A or B or total)}
\label{fig:figS3}
\end{figure}

\begin{figure}[t]
\includegraphics[width=3in]{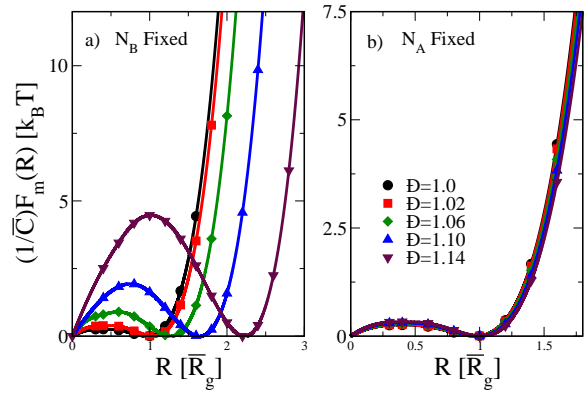}
\caption{Micelle free energy vs. micelle radius for solvophobic
fraction (a) $N_B$ {\it{Fixed}} and (b) $N_A$ {\it{Fixed}}. Symbols show SCF results, lines
a fit to the fourth order polynomial $\sum_{n=1}^4 a_n x^n$. $\DJ$ refers to dispersity in the corresponding polydisperse part of the chain (A or B)}
\label{fig:figS4}
\end{figure}

In the following, we first show the data  for \revision{the chain length distribution in the micelles compared to the bulk ( for the systems with $N_B$ {\it{Fixed}} and $N_A$ {\it{Fixed}} Fig. \ref{fig:figS2}) then the average length of the solvophobic part of the chain in the micelle for all the three systems investigated in this work (Fig.\ \ref{fig:figS3}) and then data for the
micelle free energies as a function of micelle radius for the systems with $N_B$ {\it{Fixed}} and $N_A$ {\it{Fixed}} (Fig.\ \ref{fig:figS4}).  
}

\revision{
When length of the solvophobic part of the chain is fixed, it is seen that the chain length distribution in the micelle overlaps with that in the bulk ((Fig. \ref{fig:figS2}(right)). In contrast, at higher dispersities, chain length distribution in the micelle deviates from that in the bulk when solvophilic part of the chain is fixed ((Fig. \ref{fig:figS2}(left)). 
Interestingly, at a given \DJ, there are more solvophobic segments in the micelle formed by the system with $N_B$ {\it{Fixed}} when compared to that of $f=0.5$ (Fig. \ref{fig:figS3}). Consistent with the above results, we note from Fig.\ \ref{fig:figS4}, that the effect of polymer chain dispersity on micelle free energy $F_m$ is only prominent when the solvophobic part of the chain is polydisperse in nature.}

\end{document}